\documentclass[aps,prl,twocolumn,groupedaddress,footinbib]{revtex4}
 
\usepackage{amsthm}
\usepackage{amssymb}
\usepackage{amsfonts}
\usepackage{graphicx}
\usepackage{amsmath}
\usepackage[sort&compress]{natbib}

\newcommand{\Tr}{\textrm{Tr}}

\newcommand{\bra}[1]
{
\langle #1|
}

\newcommand{\ket}[1]
{
| #1 \rangle
}

\newcommand{\showimage}[3]
{
\begin{figure}[htbp]
\begin{center}
	\includegraphics[width=#2 cm]{{#1}.jpg}
	\caption{#3}
	\label{fig:#1}
\end{center}
\end{figure}
}

\begin{document}
\title{Discontinuous quantum evolutions in the presence of closed timelike curves}

\author{Richard DeJonghe}
 \author{Kimberly Frey}
 \author{Tom Imbo}

\email[electronic address: imbo@uic.edu]{}
\affiliation{Department of Physics, University of Illinois at Chicago, 845 W. Taylor St., Chicago, IL 60607}

\begin{abstract}
We consider Deutsch's computational model of a quantum system evolving in a spacetime containing closed timelike curves.  Although it is known that this model predicts non-linear and non-unitary evolutions of the system, we demonstrate that it also gives rise to evolutions which are a discontinuous function of the input state.  These discontinuities persist for the most natural modifications of Deutsch's approach. \end{abstract}



\maketitle

Many solutions of Einstein's equations, such as G\"odel's rotating universe~\cite{Godel}, ``warp drive'' spacetimes~\cite{Alcubierre}, wormholes~\cite{Morris, Morris2}, rotating cylinders of dust~\cite{vStockum} or light~\cite{Mallet}, and the Kerr black hole~\cite{Carter}, contain closed timelike curves (CTC's).  Some physicists, most notably Hawking, find solutions (as in~[1--6]) in which the CTC's are not hidden behind event horizons objectionable on philosophical grounds due to the paradoxes which may result.  Although Hawking has proven~\cite{Hawking} that CTC's cannot be created in a bounded region in a singularity free spacetime satisfying the weak energy condition (WEC),  and has conjectured that even in spacetimes containing quantum fields (which notably do \emph{not} satisfy the WEC) no CTC's may appear at all, this famous Chronology Protection Conjecture remains unproven.  It may yet be shown that the universe is not, in fact, safe for historians.

\vspace{-.0725 in}

Leaving aside the question as to whether or not time travel is possible, a number of theoretical approaches have been developed to examine quantum systems in the presence of CTC's.  The basic strategy is to confine the closed timelike curves to a bounded region of spacetime and consider the evolution of a system from the asymptotic past, through the region containing CTC's, into the asymptotic future.  It has been established \cite{Hartle, Thorne, Deutsch, Hawking95} (see also \cite{Anderson, Fewster}) that approaches following this prescription are generally not unitary~\cite{fn1}.  
Deutsch's simple and elegant model which utilizes the methods of the quantum theory of computation is no exception~\cite{Deutsch}.  In this approach, which will be the focus of this letter, pure states can evolve into mixed states as they traverse the region containing CTC's, signaling a loss of unitarity (similar to the case of black hole evaporation~\cite{Hawking76}).  However, standard quantum mechanics is preserved in the asymptotic past and future, and transition probabilities can be calculated in the normal fashion in any region.  In addition to the loss of unitarity, other odd effects such as non-linear and non-contractive evolutions~\cite{Deutsch, Cassidy, Bacon} appear in Deutsch's model.  Moreover, this model allows for the exact cloning~\cite{Imbo,Lopata} and perfect distinguishability~\cite{Brun2} of non-orthogonal states, as well as the possibility of solving NP-complete problems in polynomial time~\cite{Bacon, Brun1}.  Yet these are not the only peculiar features --- in particular, we will show that for certain choices of dynamics, the evolution through the region containing CTC's is actually a \emph{discontinuous} function of the initial state.  Hence, for all practical purposes the model loses its predictive power in the vicinity of those initial states which are associated with a discontinuity.  In the concluding remarks we discuss possible reactions to our result.

\vspace{.05 in}

\emph{\textbf{The Deutsch Model} ---} We consider two quantum systems evolving in a universe which contains CTC's in a bounded region (called the \emph{chronology violating region}) and in which the surrounding spacetime is foliable into a set of spacelike hypersurfaces.  One system (the time traveler) follows a CTC and is confined to the chronology violating region.  The other system (the non-time traveler) originates in the asymptotic past, traverses the chronology violating region without following any CTC's, and proceeds into the asymptotic future.  Following Deutsch, the spatial evolutions of these systems are taken to be classical so that only internal degrees of freedom are treated quantum mechanically, with the associated Hilbert spaces assumed to be finite-dimensional.  Furthermore, any effect of the systems on the spacetime metric is neglected.  

A schematic diagram of this scenario is shown in Fig.~\ref{fig:canonical}.  In what follows, we denote by $\mathcal{H}_1$ and $\mathcal{H}_2$ the Hilbert spaces associated with the non-time traveling and time traveling systems, respectively, and denote by $\mathcal{D}_{\mathcal{H}}$ the set of all density operators on a given Hilbert space~$\mathcal{H}$~\cite{fn2}.
The density operators themselves (which represent the states of the systems) will be denoted by greek letters~---~for example, $\rho \in \mathcal{D}_{\mathcal{H}_1}$ will refer to the state of the non-time traveler upon entering the chronology violating region, and $\sigma \in \mathcal{D}_{\mathcal{H}_2}$ will refer to the time traveling system upon initial encounter with the non-time traveler.  According to Deutsch's prescription, the composite state~$\Omega$ of the two systems at the moment of their first encounter is taken to be uncorrelated so that $\Omega = \rho \otimes \sigma$.  

Furthermore, in this model all interactions between the two systems are assumed to be short-ranged (relative to the size of the CTC), and are represented by a single unitary gate~$U$.  After the interaction, the state of the composite system is given by $\widehat{\Omega} = U(\rho \otimes \sigma)U^{\dagger}$.  The non-time traveler then leaves the chronology violating region, and its final state~$\hat{\rho}$ is obtained by taking the partial trace over the degrees of freedom associated with the time traveler~\cite{fn3}:
\begin{equation}
\label{eq:out}
\hat{\rho} = \Tr_2(U (\rho \otimes \sigma) U^\dag).
\end{equation}
Similarly, the final state of the time traveler is given by
\begin{equation}
\label{eq:out2}
\hat{\sigma} = \Tr_1(U (\rho \otimes \sigma) U^\dag).
\end{equation}
Moreover, since the time traveler follows a closed timelike curve, traveling back in time to again meet the non-time traveler, kinematical consistency requires $\sigma = \hat{\sigma}$.  That is,  
\begin{equation}
\label{eq:cc}
\sigma = \Tr_1(U (\rho \otimes \sigma) U^\dag).  
\end{equation}
The set of all solutions to this consistency condition (for a given $U$ and $\rho$) will be denoted by $Q_{\scriptscriptstyle{U}}(\rho)$ .  Explicitly, 
\begin{equation}
\label{eq:ctts}
Q_{\scriptscriptstyle{U}}(\rho) \equiv \{ \sigma \in \mathcal{D}_{\mathcal{H}_2} \ | \ \sigma = \Tr_1(U (\rho \otimes \sigma) U^\dag) \ \}.
\end{equation}

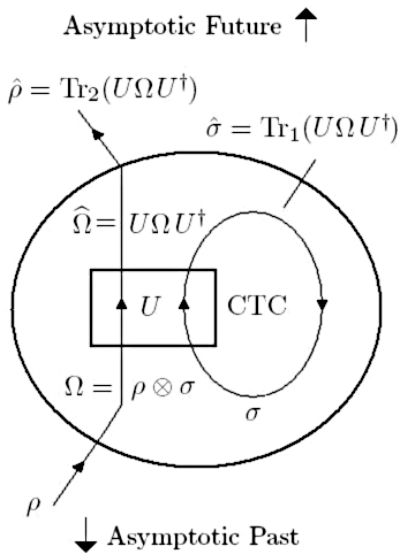
\showimage{canonical}{5.5}{A schematic diagram of the Deutsch model}

Deutsch has shown that $Q_{\scriptscriptstyle{U}}(\rho)$ is always non-empty. (This follows from Schauder's fixed point theorem \cite{Nielsen}.)  However, when $Q_{\scriptscriptstyle{U}}(\rho)$ has more than one element the model may not be deterministic  --- that is, the final state of the non-time traveler is not in general well-defined since Eq.~(\ref{eq:out}) may yield a different $\hat{\rho}$ for each $\sigma$ satisfying Eq.~(\ref{eq:cc}).  This ambiguity can be resolved --- and hence the system made deterministic --- if we have recourse to some physically motivated principle which chooses (for each $U$ and $\rho$) a unique $\sigma \in Q_{\scriptscriptstyle{U}}(\rho)$.  In such a case, the state associated with the time traveling system becomes a function of $U$ and $\rho$ which we denote by $\sigma_{\scriptscriptstyle{U}}(\rho)$.

One point of contention in the literature is related to the fact that there does not exist an obvious physical principle for choosing a unique consistent time traveling state for each $U$ and $\rho$ --- that is, for choosing the function $\sigma_{\scriptscriptstyle{U}}(\rho)$.  In an attempt to avoid the creation of information \emph{ex nihilo}, Deutsch introduces the \emph{evolutionary principle} which selects the density operator  $\sigma \in Q_{\scriptscriptstyle{U}}(\rho)$ of maximum von~Neumann entropy $S(\sigma) = - \Tr (\sigma \ln \sigma)$.  Politzer~\cite{Politzer} suggests that an equally valid principle is one which selects from $Q_{\scriptscriptstyle{U}}(\rho)$ the density operator $\sigma$ which minimizes~$S$~---~a little thought, however, shows that $S$ need not possess a unique minimum.  Although Bacon~\cite{Bacon} suggests that Deutsch's evolutionary principle also fails to choose a unique $\sigma$, the following argument shows that this is not the case.  The entropy~$S$ is a concave function of $\sigma$, which means that it satisfies (for $\sigma_a, \sigma_b \in \mathcal{D}_{\mathcal{H}_2}$ and $0 < \lambda < 1$)
\begin{equation}
\lambda S(\sigma_a) + (1- \lambda)S(\sigma_b) \leq S( \lambda \sigma_a + (1 - \lambda) \sigma_b). 
\end{equation}
Moreover, $S$ is \emph{strictly} concave as the equality holds if and only if $\sigma_a = \sigma_b$~\cite{Nielsen}.  Since $Q_{\scriptscriptstyle{U}}(\rho)$ is easily seen to be a convex subset of $\mathcal{D}_{\mathcal{H}_2}$, and any strictly concave function on a convex set has a unique maximum, it follows that there exists a unique $\sigma \in Q_{\scriptscriptstyle{U}}(\rho)$ of maximum entropy.  Of course, there may be other principles which are at least as well-motivated as the evolutionary principle and which also yield a unique choice for~$\sigma$. 

Another contentious point concerns Deutsch's assumption that the  ``initial'' state $\Omega$ of the composite system is a simple tensor product.  Politzer~\cite{Politzer} argues that this is unjustified since the systems have interacted in the causal past of the time traveler, and suggests that a more natural assumption would be to require only that $\rho = \Tr_2(\Omega)$.  However, (as he notes) this less stringent condition reduces to $\Omega = \rho \otimes \sigma$ when $\rho$ is a pure state.  

We will not be concerned with resolving either of these points of contention since our results are independent of how these issues are settled, as we will show in the following section.

\vspace{.1 in}

\emph{\textbf{Discontinuities} ---} We now demonstrate that there exist gates $U$ for which $\hat{\rho}$ is a discontinuous function of $\rho$, regardless of the choice made for $\sigma_{\scriptscriptstyle{U}}(\rho)$.    
We begin with two useful definitions.  For a given $U$, if there exists no choice of $\sigma_{\scriptscriptstyle{U}}(\rho)$ which is a continuous function of $\rho$, then we will call $U$ an \emph{ephemerally discontinuous gate}, while if there exists no choice of $\sigma_{\scriptscriptstyle{U}}(\rho)$ such that $\hat{\rho}$ is a continuous function of $\rho$, we will call $U$ a \emph{physically discontinuous gate}.  (Note that a physically discontinuous gate is ephemerally discontinuous, but the converse is not necessarily true.)  

An example of a physically discontinuous gate $U$ which acts on the Hilbert space of a three qubit system can be constructed from a controlled swap gate followed by two controlled CNOT gates; more specifically (see~Fig.~\ref{fig:discgate}), $U = \ket{000}\bra{100} + \ket{001}\bra{001} + \ket{010}\bra{011} + \ket{011}\bra{010} + \ket{100}\bra{000} + \ket{101}\bra{110} + \ket{110}\bra{101} + \ket{111}\bra{111}$.
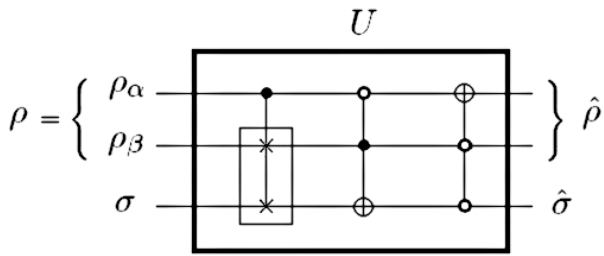
\showimage{discgate}{8.5}{A schematic diagram of a physically discontinuous gate.  The solid dots indicate a controlled gate activated for the $\ket{1}$ state, while the empty dots indicate a controlled gate activated for the $\ket{0}$ state.  The boxed $\times$'s represent a swap gate on two qubits, while the $\oplus$'s represent NOT gates.}
Here, the non-time traveler is a two qubit system, while the time traveler is a single qubit.  For notational convenience we let $\mathcal{H}_1 = \mathcal{H} \otimes \mathcal{H}$ and $\mathcal{H}_2 = \mathcal{H}$, where $\dim(\mathcal{H})~=~2$ and we have taken $ \{\ket{0}, \ket{1} \} $ as an orthonormal basis for~$\mathcal{H}$.  Also, we only consider initial states of the non-time traveler of the form $\rho = \rho_{\alpha} \otimes \rho_{\beta}$, since this restricted set of states suffices to show that discontinuous evolutions exist.  (The initial state of the composite system, per Deutsch's prescription, is given by $\Omega = \rho \otimes \sigma$, with $\rho$ as above.)
The consistency condition (Eq.~\ref{eq:out2}) becomes  
\begin{eqnarray}
(\sigma)_{11} & = & (\rho_{\beta})_{11} + (\rho_{\alpha})_{11} \Big[ 2 (\rho_{\beta})_{11} - 1 \Big] \Big[(\sigma)_{11} - 1 \Big] \\
{} & {} & {} \nonumber \\
(\sigma)_{12} & = & (\rho_{\beta})_{11} (\rho_{\alpha})_{21}(\sigma)_{12} + (\rho_{\alpha})_{11} (\rho_{\beta})_{22} (\sigma)_{21} \nonumber \\
{} & {} & + \, (\rho_{\beta})_{12} \Big[ (\sigma)_{22} (\rho_{\alpha})_{22} + (\rho_{\alpha})_{12} (\sigma)_{11} \Big] ,  
\end{eqnarray}
where $(\star)_{ij}$ denotes the ``$i,j$'' component of the appropriate density matrix in the aforementioned basis.  In what follows, we demonstrate that $U$ is ephemerally discontinuous by considering the sets $Q_{\scriptscriptstyle{U}}(\rho)$ for three distinct initial states ($\rho = \rho^{A}, \rho^{B}, \rho^{C}$) which are infinitesimally close to each other, and then using these to show that there is no choice of $\sigma_{\scriptscriptstyle{U}}(\rho)$ which is continuous in the vicinity of $\rho^{B}$.  This analysis is then further used to establish that $U$ is physically discontinuous.

We begin with $\rho^{B} = \rho^{B}_{\alpha} \otimes \rho^{B}_{\beta} = \ket{0} \bra{0} \otimes \ket{0} \bra{0}$.  For this initial state, the consistency condition gives that $Q_{\scriptscriptstyle{U}}(\rho^B) = \{ \sigma \in \mathcal{D}_{\mathcal{H}_2} \ | \ (\sigma)_{12} = 0 \}$.  (That is, any $\sigma$ along the ``$z$-axis'' of the Bloch ball $\mathcal{D}_{\mathcal{H}_2}$ is consistent.)  Now, letting $0 < \epsilon < 1$, we take $\rho^{A} = \rho^{A}_{\alpha} \otimes \rho^{A}_{\beta}$ to satisfy $(\rho^{A}_{\alpha})_{11} = 1$ and ${(\rho^{A}_{\beta})_{11} = 1 - \epsilon}$.  Then, for any such~$\epsilon$, $Q_{\scriptscriptstyle{U}}(\rho^A) = \{ \, \frac{1}{2} (\ket{0}\bra{0} + \ket{1}\bra{1}) \, \}$.  (That is, the only consistent $\sigma \in \mathcal{D}_{\mathcal{H}_2}$ is the ``chaotic state'' $\sigma^{A}$ at the center of the Bloch ball.)  Finally, taking $\rho^{C} = \rho^{C}_{\alpha} \otimes \rho^{C}_{\beta}$ to satisfy $(\rho^{C}_{\alpha})_{11} = 1 - \epsilon$ and $(\rho^{C}_{\beta})_{11} = 1$, we have that $Q_{\scriptscriptstyle{U}}(\rho^C) = \{ \, \ket{0}\bra{0} \, \}$ (independent of $\epsilon$), so that again there is only one consistent time traveling state $\sigma^{C}$ (the ``north pole'' of the Bloch ball).  

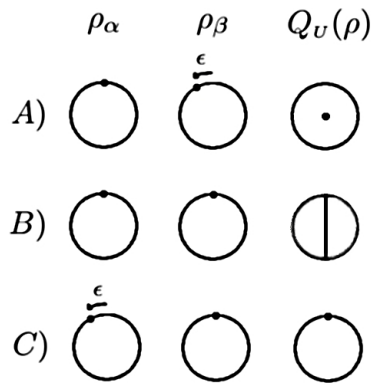
\showimage{3qubitdiscexamp_v2_rev}{5.5}{Bloch ball representation of the relevant sets of density operators for the non-time traveling system (first two columns) and the time traveling system (third column) for the example in the text.  The discs represent slices of the Bloch ball in the ``$xz$-plane'', and the top of each disc corresponds to the state $\ket{0}\bra{0}$.  The dots in the first two columns represent choices for $\rho_{\alpha}^{J}$ and $\rho_{\beta}^{J}$ ($J \in \{ A, B, C \}$), while the dots and the line in the third column represent the sets $Q_{\scriptscriptstyle{U}}(\rho^J)$.}

The above results are illustrated in Fig.~\ref{fig:3qubitdiscexamp_v2_rev}.  Since as $\epsilon \rightarrow 0$ the states  $\rho^{A}$, $\rho^{B}$, and $\rho^{C}$ become arbitrarily close, we see that although $\sigma^{A}$ and $\sigma^{C}$ are both elements of the set $Q_{\scriptscriptstyle{U}}(\rho^B)$, there is no way to choose a consistent time traveling state from $Q_{\scriptscriptstyle{U}}(\rho^B)$ such that $\sigma_{\scriptscriptstyle{U}}(\rho)$ is continuous.  Thus, a discontinuity exists about the state $\rho^{B}$ regardless of the principle used to choose $\sigma_{\scriptscriptstyle{U}}(\rho)$, showing that $U$ is an ephemerally discontinuous gate.

To see that this discontinuity is physical, consider (see~Eq.~\ref{eq:out}) 
\begin{equation} 
(\hat{\rho})_{11}  = (\rho_{\beta})_{11}  \Big[ (\rho_{\alpha})_{11} + (\sigma)_{11} - 2 (\rho_{\alpha})_{11} (\sigma)_{11} \Big]. 
\end{equation}
For $\rho = \rho^{A}$, we have that $(\hat{\rho}^{A})_{11}  = \frac{1}{2}(1 - \epsilon)$, while for $\rho = \rho^{C}$ we have that $(\hat{\rho}^{C})_{11}  = \epsilon$.  Thus, in the limit $\epsilon \rightarrow 0$ there is a discontinuous jump in the output state of the non-time traveler, showing that $U$ is physically discontinuous.

Note that in the above example we can take both $\rho^{A}$ and $\rho^{C}$ to be pure states for any value of $\epsilon$.   For all such $\rho$, any choice of the initial state $\Omega$ of the composite system consistent with Politzer's condition $\rho = \Tr_2(\Omega)$ will necessarily be uncorrelated.  Thus, in this case Politzer's generalization reduces to Deutsch's assumption ($\Omega = \rho \otimes \sigma$), so that relaxing the latter in this manner cannot keep the discontinuity from appearing.

The gate discussed above falls naturally into a larger class of unitary operators for this three qubit system --- namely those which permute the basis vectors of the Hilbert space $\mathcal{H}_1 \otimes \mathcal{H}_2$.  We have analyzed the majority of these $8!$ gates and found that almost half of those considered are physically discontinuous (while approximately two-thirds are ephemerally discontinuous).  It is quite remarkable that the discontinuous gates are ubiquitous within such a simple class of examples.  However, despite these findings, the totality of three qubit gates exhibiting discontinuities may yet be a set of measure zero in the space of all unitary operators for this system.

\emph{\textbf{Discussion} ---} We have considered Deutsch's model of a non-time traveling system interacting with a time traveler confined to a bounded region, and have demonstrated that the state of the non-time traveler in the asymptotic future can be a discontinuous function of the state in the asymptotic past.  Furthermore, we have demonstrated that these discontinuities occur independent of the method of choosing a unique consistent time traveling state, as well as independent of whether Deutsch's assumption regarding the initial composite state or Politzer's generalization is used.

Given the phenomenon of discontinuous evolutions within the Deutsch model, we note several possible reactions. 

\vspace{.02 in}
\noindent
(1)  Question the assumptions upon which Deutsch's model is based.  However, relaxing the two most obvious of these, as stated in the previous paragraph, does not provide any respite.  Thus, the only remaining natural assumptions to be questioned are that (a)~the spatial degrees of freedom can be treated classically, (b)~the effect of the systems on the surrounding spacetime can be neglected, and finally that (c)~a quantum mechanical (as opposed to field-theoretic) model captures the relevant dynamics.  Although over-idealizations can indeed lead to apparent discontinuities, none of (a)--(c) above seems obviously responsible for the discontinuous behavior in Deutsch's model.  In particular, it is diffcult to believe that there is no imaginable configuration utilizing a discontinuous gate for which these approximations are sufficiently justifed.

\vspace{.04 in}
\noindent
(2)  Accept the assumptions of the Deutsch model, but further assume that nature either does not utilize those gates which are physically discontinuous, or does not allow initial states of the non-time traveler which are near a discontinuity.  (Analagous tactics have been considered in the classical case as a way of avoiding the grandfather paradox \cite{Deutsch}.)  However, this solution is somewhat \emph{ad hoc} and inelegant.  In addition, placing such restrictions on initial states and/or gates sacrifices one of the great strengths of Deutsch's approach which purports to provide a viable model for \emph{any} set of initial conditions and \emph{any} dynamics.

\vspace{.04 in}
\noindent 
(3)  Accept that the Deutsch model is correct as writ, but interpret the existence of discontinuous evolutions as evidence that CTC's are unphysical.  

\vspace{.04 in}
\noindent 
(4)  Acknowledge that quantum mechanics in the presence of CTC's is sufficiently strange that the existence of these discontinuities is a fitting physical consequence. 

\vspace{.04 in}
Further study will be required not only to adequately address these reactions, but also to answer other interesting questions raised by our results, such as:  What are the exact properties of the gates which give rise to such peculiar evolutions?  For any such gate, how are the points at which the evolution is discontinuous distributed in the space of initial states?
Do these discontinuities occur in other approaches to quantum systems in the presence of CTC's?  
Regardless, it is clear that discontinuous evolutions are an unavoidable feature of the Deutsch model, and are yet another strange and fascinating consequence of the attempt to bring together quantum mechanics and gravity. 

\vspace{-.04 in}
We thank Randall Espinoza, Nick Huggett, Paul Lopata, and Mark Mueller for many useful discussions.


\vspace{-.2 in}

\begin{thebibliography}{23}

\expandafter\ifx\csname natexlab\endcsname\relax\def\natexlab#1{#1}\fi
\expandafter\ifx\csname bibnamefont\endcsname\relax
  \def\bibnamefont#1{#1}\fi
\expandafter\ifx\csname bibfnamefont\endcsname\relax
  \def\bibfnamefont#1{#1}\fi
\expandafter\ifx\csname citenamefont\endcsname\relax
  \def\citenamefont#1{#1}\fi
\expandafter\ifx\csname url\endcsname\relax
  \def\url#1{\texttt{#1}}\fi
\expandafter\ifx\csname urlprefix\endcsname\relax\def\urlprefix{URL }\fi
\providecommand{\bibinfo}[2]{#2}
\providecommand{\eprint}[2][]{\url{#2}}

\bibitem[{\citenamefont{G\"odel}(1949)}]{Godel}
\bibinfo{author}{\bibfnamefont{K.}~\bibnamefont{G\"odel}},
  \bibinfo{journal}{Rev. Mod. Phys.} \textbf{\bibinfo{volume}{21}},
  \bibinfo{pages}{447} (\bibinfo{year}{1949}).

\bibitem[{\citenamefont{Alcubierre}(1994)}]{Alcubierre}
\bibinfo{author}{\bibfnamefont{M.}~\bibnamefont{Alcubierre}},
  \bibinfo{journal}{Class. Quant. Grav.} \textbf{\bibinfo{volume}{11}},
  \bibinfo{pages}{L73} (\bibinfo{year}{1994}).

\bibitem[{\citenamefont{Morris and Thorne}(1988)}]{Morris}
\bibinfo{author}{\bibfnamefont{M.}~\bibnamefont{Morris}} \bibnamefont{and}
  \bibinfo{author}{\bibfnamefont{K.}~\bibnamefont{Thorne}},
  \bibinfo{journal}{Am. J. Phys} \textbf{\bibinfo{volume}{56}},
  \bibinfo{pages}{395} (\bibinfo{year}{1988}).

\bibitem[{\citenamefont{Morris et~al.}(1988)\citenamefont{Morris, Thorne, and
  Yurtsever}}]{Morris2}
\bibinfo{author}{\bibfnamefont{M.}~\bibnamefont{Morris}},
  \bibinfo{author}{\bibfnamefont{K.}~\bibnamefont{Thorne}}, \bibnamefont{and}
  \bibinfo{author}{\bibfnamefont{U.}~\bibnamefont{Yurtsever}},
  \bibinfo{journal}{Phys. Rev. Lett.} \textbf{\bibinfo{volume}{61}},
  \bibinfo{pages}{1446} (\bibinfo{year}{1988}).

\bibitem[{\citenamefont{van Stockum}(1937)}]{vStockum}
\bibinfo{author}{\bibfnamefont{W.~J.} \bibnamefont{van Stockum}},
  \bibinfo{journal}{Proc. R. Soc. Edin.} \textbf{\bibinfo{volume}{57}},
  \bibinfo{pages}{135} (\bibinfo{year}{1937}).

\bibitem[{\citenamefont{Mallett}(2003)}]{Mallet}
\bibinfo{author}{\bibfnamefont{R.~L.} \bibnamefont{Mallett}},
  \bibinfo{journal}{Found. Phys.} \textbf{\bibinfo{volume}{33}},
  \bibinfo{pages}{1307} (\bibinfo{year}{2003}).

\bibitem[{\citenamefont{Carter}(1968)}]{Carter}
\bibinfo{author}{\bibfnamefont{B.}~\bibnamefont{Carter}},
  \bibinfo{journal}{Phys. Rev.} \textbf{\bibinfo{volume}{174}},
  \bibinfo{pages}{1559} (\bibinfo{year}{1968}).

\bibitem[{\citenamefont{Hawking}(1992)}]{Hawking}
\bibinfo{author}{\bibfnamefont{S.~W.} \bibnamefont{Hawking}},
  \bibinfo{journal}{Phys Rev. D} \textbf{\bibinfo{volume}{46}},
  \bibinfo{pages}{603} (\bibinfo{year}{1992}).

\bibitem[{\citenamefont{Hartle}(1994)}]{Hartle}
\bibinfo{author}{\bibfnamefont{J.}~\bibnamefont{Hartle}},
  \bibinfo{journal}{Phys. Rev. D} \textbf{\bibinfo{volume}{49}},
  \bibinfo{pages}{6543} (\bibinfo{year}{1994}).

\bibitem[{\citenamefont{Echeverria et~al.}(1991)\citenamefont{Echeverria,
  Klinkhammer, and Thorne}}]{Thorne}
\bibinfo{author}{\bibfnamefont{F.}~\bibnamefont{Echeverria}},
  \bibinfo{author}{\bibfnamefont{G.}~\bibnamefont{Klinkhammer}},
  \bibnamefont{and} \bibinfo{author}{\bibfnamefont{K.~S.}
  \bibnamefont{Thorne}}, \bibinfo{journal}{Phys Rev D}
  \textbf{\bibinfo{volume}{44}}, \bibinfo{pages}{1077} (\bibinfo{year}{1991}).

\bibitem[{\citenamefont{Deutsch}(1991)}]{Deutsch}
\bibinfo{author}{\bibfnamefont{D.}~\bibnamefont{Deutsch}},
  \bibinfo{journal}{Phys. Rev. D} \textbf{\bibinfo{volume}{44}},
  \bibinfo{pages}{3197} (\bibinfo{year}{1991}).

\bibitem[{\citenamefont{Hawking}(1995)}]{Hawking95}
\bibinfo{author}{\bibfnamefont{S.~W.} \bibnamefont{Hawking}},
  \bibinfo{journal}{Phys. Rev. D} \textbf{\bibinfo{volume}{52}},
  \bibinfo{pages}{5681} (\bibinfo{year}{1995}).

\bibitem[{\citenamefont{Anderson}(1995)}]{Anderson}
\bibinfo{author}{\bibfnamefont{A.}~\bibnamefont{Anderson}},
  \bibinfo{journal}{Phys. Rev. D} \textbf{\bibinfo{volume}{51}},
  \bibinfo{pages}{5707} (\bibinfo{year}{1995}).

\bibitem[{\citenamefont{Fewster and Wells}(1995)}]{Fewster}
\bibinfo{author}{\bibfnamefont{C.~J.} \bibnamefont{Fewster}} \bibnamefont{and}
  \bibinfo{author}{\bibfnamefont{C.}~\bibnamefont{Wells}},
  \bibinfo{journal}{Phys Rev. D} \textbf{\bibinfo{volume}{52}},
  \bibinfo{pages}{5773} (\bibinfo{year}{1995}).


\bibitem[{\citenamefont{fn1}}(2009)]{fn1}
Although this loss of unitarity can be remedied within some approaches, doing so typically leads to other problematic aspects.  For example, the models of Hartle~\cite{Hartle} and Echeverria et~al.~\cite{Thorne} restore unitarity at the cost of introducing a non-linearity so severe that it affects events in the asymptotic past.  Anderson~\cite{Anderson}, on the other hand, shows that unitarity can be restored while maintaining causality, but at the cost of introducing evolutions which do not obey the normal composition law (see~\cite{Fewster}).


\bibitem[{\citenamefont{Hawking}(1976)}]{Hawking76}
\bibinfo{author}{\bibfnamefont{S.~W.} \bibnamefont{Hawking}},
  \bibinfo{journal}{Phys Rev D.} \textbf{\bibinfo{volume}{14}},
  \bibinfo{pages}{2460} (\bibinfo{year}{1976}).

\bibitem[{\citenamefont{Cassidy}(1995)}]{Cassidy}
\bibinfo{author}{\bibfnamefont{M.}~\bibnamefont{Cassidy}},
  \bibinfo{journal}{Phys. Rev. D} \textbf{\bibinfo{volume}{52}},
  \bibinfo{pages}{5676} (\bibinfo{year}{1995}).

\bibitem[{\citenamefont{Bacon}(2004)}]{Bacon}
\bibinfo{author}{\bibfnamefont{D.}~\bibnamefont{Bacon}},
  \bibinfo{journal}{Phys. Rev. A} \textbf{\bibinfo{volume}{70}},
  \bibinfo{pages}{032309} (\bibinfo{year}{2004}).

\bibitem[{\citenamefont{Imbo and Lopata}(2009)}]{Imbo}
\bibinfo{author}{\bibfnamefont{T.}~\bibnamefont{Imbo}} \bibnamefont{and}
  \bibinfo{author}{\bibfnamefont{P.}~\bibnamefont{Lopata}},
  \bibinfo{journal}{in preparation}  (\bibinfo{year}{2009}).

\bibitem[{\citenamefont{Lopata}(2005)}]{Lopata}
\bibinfo{author}{\bibfnamefont{P.}~\bibnamefont{Lopata}}, \bibinfo{journal}{PhD
  dissertation}  (\bibinfo{year}{2005}).

\bibitem[{\citenamefont{Brun et~al.}(2009)\citenamefont{Brun, Harrington, and
  Wilde}}]{Brun2}
\bibinfo{author}{\bibfnamefont{T.~A.} \bibnamefont{Brun}},
  \bibinfo{author}{\bibfnamefont{J.}~\bibnamefont{Harrington}},
  \bibnamefont{and} \bibinfo{author}{\bibfnamefont{M.~M.} \bibnamefont{Wilde}},
  \bibinfo{journal}{Phys. Rev. Lett.} \textbf{\bibinfo{volume}{102}},
  \bibinfo{pages}{210402} (\bibinfo{year}{2009}).

\bibitem[{\citenamefont{Brun}(2003)}]{Brun1}
\bibinfo{author}{\bibfnamefont{T.~A.} \bibnamefont{Brun}},
  \bibinfo{journal}{Found. Phys. Lett.} \textbf{\bibinfo{volume}{16}},
  \bibinfo{pages}{245} (\bibinfo{year}{2003}).

\bibitem[{\citenamefont{fn2}}(2009)]{fn2}
We regard each $\mathcal{D}_{\mathcal{H}}$ as embedded in $\mathbb{R}^n$ in the standard way, where $n=(dim(\mathcal{H}))^2-1$.  The usual topology on $\mathbb{R}^n$ then induces a topology on $\mathcal{D}_{\mathcal{H}}$, which we assume throughout.


\bibitem[{\citenamefont{fn3}}(2009)]{fn3}
Deutsch considers a more general dynamics in which a finite number of systems originate in the asymptotic past, and a subset of these systems follow CTC's and interact with younger versions of themselves while a (possibly different) subset interacts with any number of other time travelers which are confined to the chronology violating region and which follow distinct CTC's.  However, as shown in \cite{Deutsch}, any such case is equivalent to a situation of the type we consider (with an appropriate choice of $U$, $\mathcal{H}_1$ and $\mathcal{H}_2$) in the sense that the final state $\hat{\rho}$ will be the same function of the initial state $\rho$.


\bibitem[{\citenamefont{Nielsen and Chuang}(2000)}]{Nielsen}
\bibinfo{author}{\bibnamefont{Nielsen}} \bibnamefont{and}
  \bibinfo{author}{\bibnamefont{Chuang}}, \emph{\bibinfo{title}{Quantum
  Computation and Quantum Information}} (\bibinfo{publisher}{Cambridge
  University Press}, \bibinfo{year}{2000}).

\bibitem[{\citenamefont{Politzer}(1994)}]{Politzer}
\bibinfo{author}{\bibfnamefont{H.}~\bibnamefont{Politzer}},
  \bibinfo{journal}{Phy. Rev. D} \textbf{\bibinfo{volume}{49}},
  \bibinfo{pages}{3981} (\bibinfo{year}{1994}).

\end{thebibliography}
\end{document}